\documentclass[amssymb,aps,floatfix,floats,preprintnumbers,shownopacs,twocolumn]{revtex4-1}
\usepackage{rcs}
\usepackage{graphicx}
\usepackage{dcolumn}
\usepackage{bm}
\usepackage[applemac]{inputenc} 
\usepackage{color}

\begin{document}
\title{Thermodynamic evidence for valley-dependent density of states in bulk bismuth}
\author{R. Kuechler$^{1\dagger}$, L. Steinke$^{1\dagger}$, R. Daou$^{1,2}$, M. Brando$^1$, K. Behnia$^3$, and F. Steglich$^1$}

\affiliation{$^1$\mbox{Max Planck Institute for Chemical Physics of Solids, D-01187 Dresden, Germany} \\
$^2$Crismat, UMR 6508, F-14050 Caen, France \\
$^3$\mbox{LPEM (UPMC-CNRS), ESPCI, F-75231, Paris, France}\\
$^\dagger$\mbox{These authors contributed equally to this work}} 
\date{\today}

\begin{abstract}

\textbf{Electron-like carriers in bismuth are described by the Dirac Hamiltonian, with a band mass becoming a thousandth of the bare electron mass along one crystalline axis~\cite{Dresselhaus71}. The existence of three anisotropic valleys offers electrons an additional degree of freedom, a subject of recent attention~\cite{Zhu11a}. Here, we map the Landau spectrum by angle-resolved magnetostriction, and quantify the carrier number in each valley: while the electron valleys keep identical spectra, they substantially differ in their density of states at the Fermi level. Thus, the electron fluid does not keep the rotational symmetry of the lattice at low temperature and high magnetic field, even in the absence of internal strain. This effect, reminiscent of the Coulomb pseudo-gap in localized electronic states, affects only electrons in the immediate vicinity of the Fermi level. It presents the most striking departure from the non-interacting picture of electrons in bulk bismuth.}

\end{abstract}

\maketitle
\begin{figure*}
\centering
  \begin{tabular}{@{}ccc@{}}
    \includegraphics{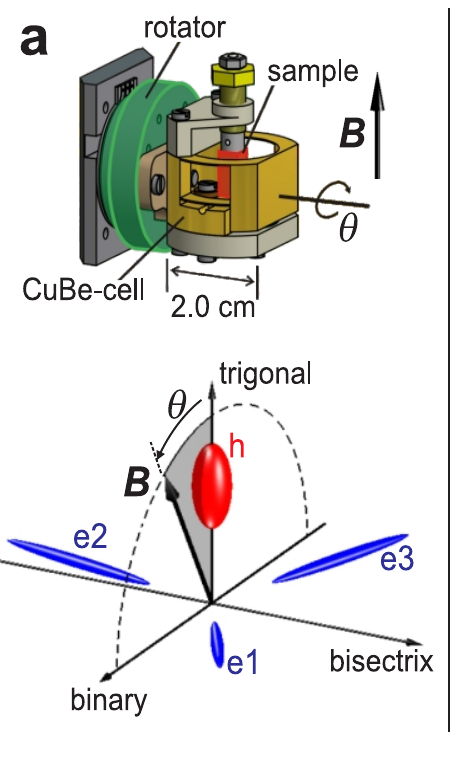} &
    \includegraphics{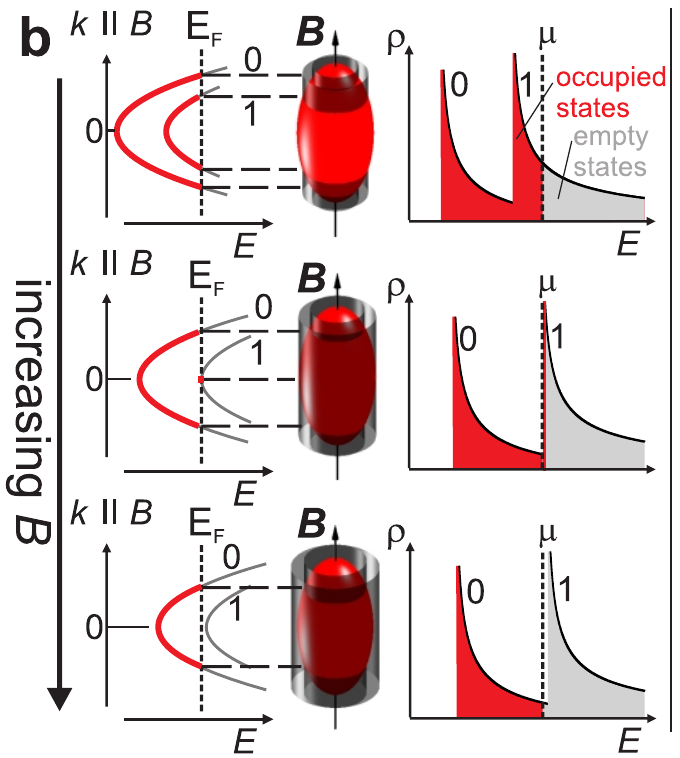} &
    \includegraphics{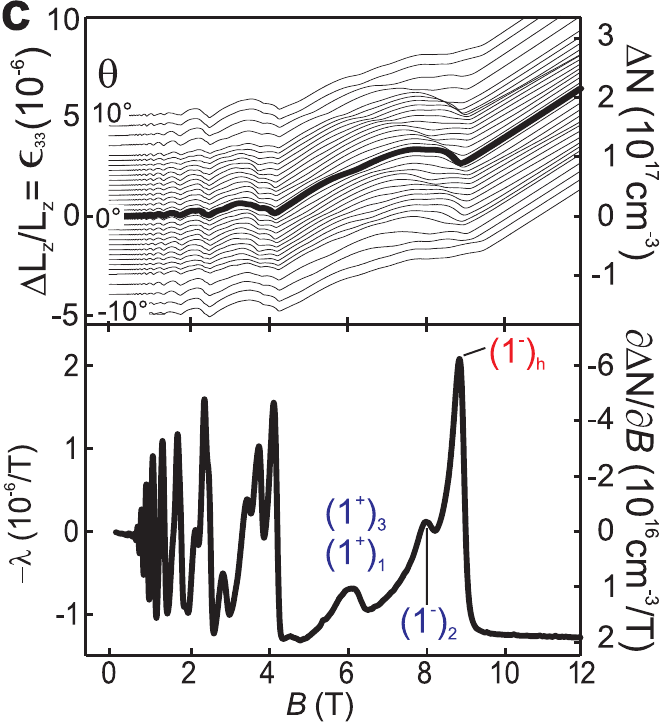}
    \end{tabular}
\caption{\textbf{Angle-dependent magnetostriction measurements.} {\bf a}, Experimental setup with the sample mounted inside a capacitive dilatometer cell attached to a piezo-rotator, to allow for {\it in situ} rotation at low temperatures. Below: magnetic field orientation relative to the Fermi surface, which in Bismuth consists of a hole pocket (red) along the trigonal axis, and three electron pockets (blue) close to the binary-bisectrix plane.
{\bf b}, Schematic dispersion $E(k\vert\vert B)$ and density of states $\rho(E)$, with illustrations showing the emptying of the second LL with increasing $B$.
{\bf c}, top: Measurement of $\epsilon_{33}(B)$ at various angles $-10^{\circ}\leq\theta\leq+10^{\circ}$, and $T = 0.025$ K. The right-hand y-axis shows the density change $\Delta N$ calculated from $\epsilon_{33}$. {\bf c}, bottom: magnetostriction coefficient $-\lambda$ derived from the $\theta=0^{\circ}$-measurement, and calculated differential density change $\partial\Delta N/\partial(B)$.}
\label{Fig1}
\end{figure*}
A naturally occurring element known since ancient times~\cite{Agricola,Garland}, the semi-metal bismuth has played an important role in the history of solid state physics~\cite{Edelman76}. Over the last five years, bismuth has attracted new attention focused on its unusual properties in the presence of a strong magnetic field. More than 120 years after the first discovery of the Nernst-Ettingshausen effect (NE) in the very same material~\cite{Nernst}, giant NE oscillations were observed in the vicinity of the quantum limit~\cite{Behnia07a}. This limit is attained when the magnetic field is strong enough to confine carriers to their lowest Landau level(s). With currently available magnetic fields, it is only accessible in solids with a carrier concentration as low as in bismuth. Theory expects a prominent role for electron interactions beyond this limit~\cite{Halperin87}. NE measurements  extended to 33 T, well above the quantum limit~\cite{Behnia07b}, found features unexpected in the single-particle picture. Torque magnetometry studies by Li \emph{et al.}~\cite{Luli08} up to 31 T showed a structure in magnetization with hysteresis suggesting a correlation-induced phase transition.

These results inspired new theoretical studies of the Landau spectrum in bismuth~\cite{Alicea09,Sharlai09}, which is remarkably complex because the electron dispersion is Dirac-like, with an additional anisotropic Zeeman term~\cite{Vecchi76,Mcclure76}. In particular, in the vicinity of the high-symmetry axis known as the trigonal axis, the magnetic field at which an electron Landau level is depleted rapidly changes with the field orientation. Recently, the angle-resolved Landau spectrum was determined by NE measurements for the whole solid angle~\cite{Zhu12,Zhu11b}, and found to be in good agreement with theoretical calculations by Fuseya~\cite{Zhu11b}, based on an earlier model by Vecchi and co-workers~\cite{Vecchi76}. Thus the non-interacting picture can successfully explain most features of the high-field phase diagram, including those previously attributed to electron interaction~\cite{Behnia07b,Luli08}. Still the unexpected hysteresis seen in the torque measurements~\cite{Luli08}, reminiscent of a first-order phase transition, remained unexplained. However, subsequent torque magnetometry studies~\cite{Fauque09} failed to reproduce this hysteresis and found no evidence of the suspected high-field phase transition.

Most recently, a new experiment by Zhu {\it et al.} refueled the suspicions of possible interaction effects in bismuth: the magnetoconductivity tensor in bismuth was shown to spontaneously lose the threefold rotational symmetry of the underlying lattice at low temperature and high magnetic field~\cite{Zhu11a}. This behavior, which is not related to the previously reported hysteresis, is again unexpected in the non-interacting picture and suggests a type of electronic order involving the valley degree of freedom and favored by electron-electron interactions. However, a possible role of internal strain, which could lift the valley degeneracy, could not be ruled out~\cite{Unfinished_bismuth}.

In the context of this lively discussion, we revisit another historical experiment on bismuth: oscillatory magnetostriction~\cite{Cha63,Cha69}. Employing a new  miniaturized capacitance dilatometer with sub-$\mathrm{\AA}$ resolution~\cite{Kuechler12}, we establish the angle-resolved Landau spectrum with exceptional clarity, in excellent agreement with theoretical calculations~\cite{Alicea09,Sharlai09,Zhu11b}, and again with no trace of hysteresis. This observation clearly excludes the existence of a thermodynamic first order phase transition at high fields and in the tilt angle range studied by Li {\it et al.}~\cite{Luli08}. More importantly, we find compelling evidence for spontaneous valley imbalance. In contrast to transport~\cite{Zhu11a}, the thermodynamic probe employed in this study directly couples to the density of states. Moreover, for the first time our experiment establishes that while two distinct valleys keep identical Landau spectra, they present a different density of states at the Fermi level. This rules out the possibility that uncontrolled strain is responsible for lifting the valley degeneracy and sets boundaries on the possible theoretical scenarios. We will argue that the situation is reminiscent of the Coulomb gap in doped semiconductors, where Coulomb interaction induces a soft gap in the density of states~\cite{Efros75}.

We measured magnetostriction with a miniaturized capacitance dilatometer~\cite{Kuechler12} illustrated in Fig.\,\ref{Fig1}a and explained in detail in the supplement.
As seen in Fig.\,\ref{Fig1}c (top panel), the magnetic field dependent length change $\Delta L_z/L_z$ measured at  $T = 0.025$ K, with the field $B$ applied along the trigonal axis, clearly shows angle-dependent oscillatory features. At low fields, $B^{-1}$-perodic quantum oscillations were observed down to fields as low as $\sim 0.2$ T~\cite{Kuechler12}, and the period of \mbox{0.15 T$^{-1}$}, corresponding to the maximum cross-section of the hole Fermi pocket, is identical to results obtained from oscillatory resistivity~\cite{Bompadre01} and oscillatory NE~\cite{Behnia07a}. 
With increasing field, the oscillations become more pronounced, leading to the last and most prominent resonance at 9 T, clearly visible on top of the monotonous background in Fig.\,\ref{Fig1}c. This field represents the quantum limit of holes.

The magnetostriction is linked to the stress dependence of the magnetization via a Maxwell relation~\cite{Chan71}. In a compensated semi-metal, bismuth in particular, magnetostriction is generated by a field-dependent variation of electron and hole concentrations~\cite{Mic82}.
The magnetostrictive strain is a linear function of the field-induced change in carrier density $\Delta N (B)~$\cite{Mic82}. In particular, when the magnetic field is applied along the trigonal axis, the longitudinal magnetostriction is simply:
\begin{equation}
\epsilon_{33} =\frac{ \Delta L_{z}}{L_{z}} = c \Delta N (B)
\label{eq2}
\end{equation}
The magnitude of the coefficient $c$ is determined by a product of components of elastic compliance and deformation potential tensors. The elastic compliance links elastic energy to strain, and the deformation potential tensor quantifies the strain required for shifting a band edge. (See the supplement for a more detailed discussion). The magnetostriction coefficient $\lambda_{z}=d\epsilon_{33}/dB$ thus represents the derivative of the carrier number with respect to the magnetic field:
\begin{equation}
\lambda_{z} = c \frac{dN}{dB}.
\label{lambda}
\end{equation}

The applied magnetic field further quantizes the quasiparticles energies into highly degenerate Landau levels (LLs) which occupy discrete energy levels depending on the cyclotron energy perpendicular to the magnetic field direction \cite{Shoenberg88}. As the cyclotron energy increases with increasing field, the higher energy LLs become depopulated. Above the quantum limit of 9\,T the quasiparticles are confined to the lowest LL. The density of states (DOS) has a series of singularities $\sim1/\sqrt{E-E_i}$, as sketched in Fig.~\ref{Fig1}b.

The spikes seen in $-\lambda$ (Fig. \ref{Fig1}c, bottom panel) are therefore caused by the sudden changes in the carrier number whenever a LL is evacuated and they reflect the asymmetric singularities in the DOS. According to calculations~\cite{Mic82,Zhu11b}, the change in the total carrier number (of both holes and electrons) is much more drastic for the evacuation of a hole than an electron LL.  In agreement with the implications of Eq.~\ref{eq2}, we find that hole peaks in our magnetostriction data (Fig. \ref{Fig1}c) are significantly more prominent than electron peaks.

Approaching the quantum limit, only a few levels remain occupied, and an increasing fraction of the total electrons are concerned by each LL evacuation. The change in the total DOS therefore becomes more significant. Since the electric charge is to be conserved, in a single-band metal the carrier concentration remains fixed and only the chemical potential shifts after each evacuation, to satisfy the change in DOS. In a compensated system like bismuth, the situation is different. While the total number of holes should match the total number of electrons, a simultaneous change in both hole and electron concentrations can take place if it reduces the free energy. Sweeping the magnetic field thus leads to a change $\Delta N$ in carrier concentration, and a corresponding shift in band edges. The cost in the elastic free energy is compensated by the decrease in the overlap between conduction and valence bands~\cite{Keyes67}.

To establish  the angle-dependent LL spectrum for electrons and holes, we measured the magnetostriction at various tilt angles $\theta$ of the field in the trigonal-binary plane (see Fig.~\ref{Fig1}a). Oscillatory features from the subsequent depletion of Landau levels, already visible in the raw $\Delta L/L$ data in Fig.\,\ref{Fig1}c, are clearly discernible as resonances in the derived magnetostriction coefficient $\lambda$. 
\begin{figure}
\includegraphics{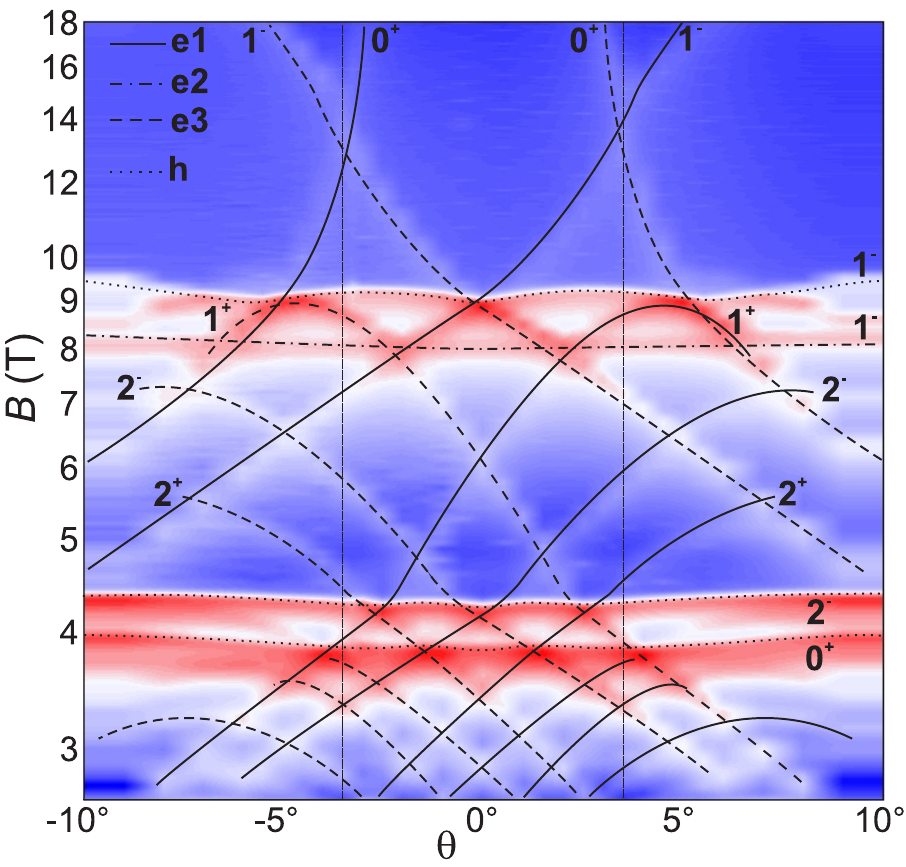}
\caption{\textbf{High-field Landau spectrum in the $B-\theta$ plane.} Color map plot of the measured magnetostriction coefficient $-\lambda$ versus tilt angle $\theta$ and magnetic field $B$. The color coding is arbitrary. The black lines are guides to the eye, linking the observed peaks to corresponding electron and hole pockets of the Fermi surface. Additional labels indicate the LL and spin index. The vertical lines serve as a reference for Fig.\,\ref{Fig3}b.}
\label{Fig2}
\end{figure}
Fig.~\ref{Fig2} shows a color map plot of $-\lambda$ vs. $B$ and $\theta$, obtained from the angle-dependent measurements in Fig.\,\ref{Fig1}c. The black lines serve as guides to the eye to highlight the measured resonances in $-\lambda$, and the different line styles relate the resonant features to the three electron pockets or the hole pocket of the bismuth Fermi surface. The additional labels $i^{\pm}$ next to the lines indicate the LL index $(i)$ and spin $(\pm)$. This is the first time that the smaller peaks associated with electron LLs in bismuth can be resolved in a magnetostriction measurement. The peak positions for the electron LLs are in good agreement with theoretical calculations by Alicea and Balents~\cite{Alicea09}, whereas the hole spectrum is better described by the calculations by Y. Fuseya in Ref.~\cite{Zhu11b}. The horizontal line at 8 T in Fig.~\ref{Fig2} is attributed to the $1^-$-resonance of electron pocket e2 (cf. Fig.~\ref{Fig1}a), which even at $\theta=0^{\circ}$ appears at a lower field than the corresponding peaks of the other two electron pockets. We explain this mismatch by a slight misalignment, where the field at $\theta=0^{\circ}$ is not perfectly aligned with the trigonal axis, but tilted towards the bisectrix axis by an estimated $-1.4^{\circ}$.

Our data do not show any of the additional, theoretically unexpected, lines observed in NE measurements at 13.3, 22.3, and 30.8 T~\cite{Behnia07b,Yang10}. According to a recent interpretation, these anomalies were caused by the presence of a secondary twinned crystal~\cite{Zhu12}. We have no reason to assume that our bismuth crystal is completely free of such defects, but our Landau spectrum does not show the associated anomalies, probably due to the small volume fraction occupied by secondary twins. 
Since magnetostriction is a bulk measurement dominated by the principal crystal domain, the presence of such minority domains can likely not be detected. Moreover, we did not observe any hysteresis in contrast to torque measurements~\cite{Luli08}, despite carefully looking for it at similar angles and magnetic fields (see Fig. S2 {b} in the supplementary information). Together with the torque magnetometry results in Ref.~\cite{Fauque09}, this is strong evidence against a possible first-order phase transition inferred from the previous work by Li {\it et al.}~\cite{Luli08}. In summary, our results confirm the validity of the one-particle picture in describing the complex angle-resolved Landau spectrum of bismuth near the trigonal axis, a conclusion shared by a recent study of angle-resolved NE effect in the whole solid angle~\cite{Zhu12}.

\begin{figure}
\includegraphics{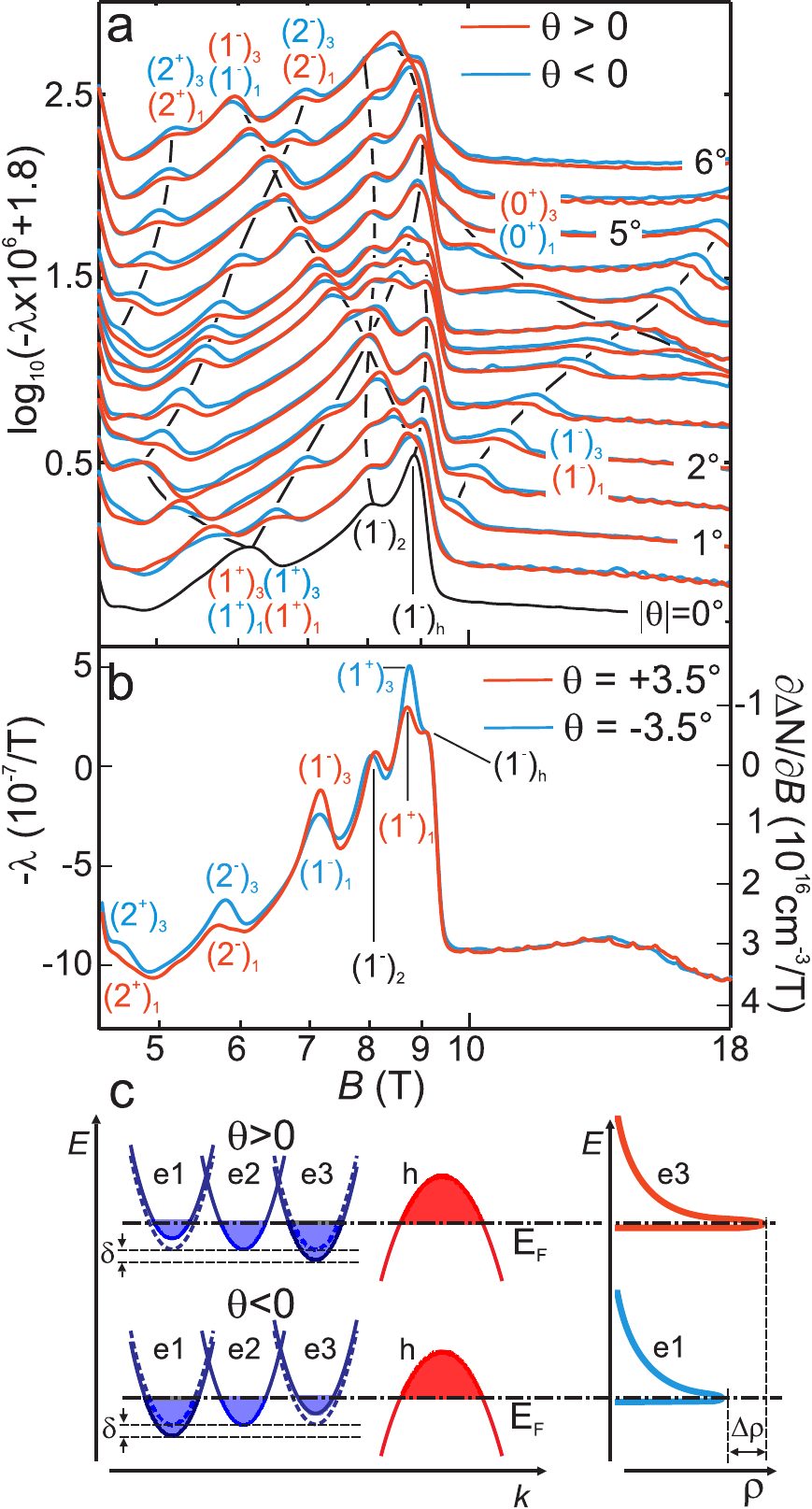}
\caption{\textbf{Valley-dependent density of states.} {\bf a}, waterfall plot of $\log_{10}(-\lambda\times10^6+1.8)$ vs. $B$, comparing measurements at positive (orange) and negative (blue) tilt angles $\theta$. Peaks are labeled with their valley index, LL and spin quantum numbers.
{\bf b}, comparison of $-\lambda(B)$ and $\partial\Delta N/\partial(B)$ at $\theta=\pm 3.5^{\circ}$ (vertical lines in Fig.\ref{Fig2}) on a linear scale.
{\bf c}, schematic angle-dependent electron and hole dispersion $E(k)$ and a sketch comparing the density of states $\rho(E)$ for two different electron pockets with respectively equal Fermi surface crossections at $\pm \theta$.}
\label{Fig3}
\end{figure}

The most important result of this study is presented in Fig.~\ref{Fig3}, in which $B$-dependent measurements of $-\lambda$ at positive (orange) and negative (blue) tilt angles $\theta$ are plotted side by side. As illustrated in panel c, the band edges for the hole pocket or electron valley 2 are insensitive to this tilt direction, and the corresponding peaks should appear symmetrically at the same fields, independent of the sign of $\theta$. In contrast, the other resonances associated with the two electron valleys 3 (or 1) are expected to shift upward (or downward) for positive (or negative) $\theta$, and measurements at opposite signs of $\theta$ can be used to directly compare these two valleys. We find that all peaks associated with electron valley 3 are significantly larger than the peaks from electron valley 1. For example, In Fig.~\ref{Fig3} (panel b) measurements at $\pm3.5^{\circ}$ are compared on a linear scale, showing a difference of \mbox{$\approx 20$ \%} of the total peak height between valleys 1 and 3. As expected, the monotonous background as well as all other resonances originating from electron valley 2 or the hole pocket produce identical values for positive and negative $\theta$. We can therefore exclude instrumental effects, like a general angle-dependence of the dilatometer sensitivity, as the possible cause for the observed anisotropy. Also a slight rotation of the sample in the binary-bisectrix plane can be ruled out considering the near-perfect mirror-symmetry of the peak positions (cf. discussion in the supplement). The peaks in $-\lambda$, corresponding to the differential density change $-\partial\Delta N/\partial B$, reflect the singularities in the DOS whenever a LL passes the Fermi level. Our observation of different peak heights for different valleys can therefore be interpreted as a valley-dependent density of states close to the depletion of a Landau level. This is the first thermodynamic evidence for the loss of three-fold symmetry by the electron fluid at low temperature.

A previous study of magnetoconductivity in bulk bismuth with a magnetic field rotating in the binary-bisectrix plane also reported a spontaneous loss of threefold symmetry in the magnetoresistance data~\cite{Zhu11a}, which could be a related effect. In the following, we discuss various scenarios that could lead to these observations. The simplest possible explanation is experimental. Symmetry breaking may arise due to frozen-in strains and/or the sample can become asymmetrically strained upon cooling. Uniaxial stress is known to drastically imbalance the valley population in bismuth~\cite{Brandt80}, so could the observed loss of threefold symmetry simply be a consequence of uncontrolled strain, as suggested in Ref.~\cite{Unfinished_bismuth}? Our experimental observation clearly rules out the possible role of uncontrolled strain. If valley degeneracy were lifted by strain, an unequal occupation of the electron valleys would be observed. However, we find that not only the amplitude of the background but also the peak positions remain almost perfectly symmetric between positive and negative angles. This means that the two valleys remain identical in size, but differ only in their DOS at the Fermi level.

Another obvious cause of a broken threefold symmetry would be secondary twinned crystals, which already led to the puzzling observation of additional lines in the angle-dependent NE spectrum~\cite{Behnia07b,Yang10,Zhu12}. However, our magnetostriction measurements do not show any such features. We therefore conclude that the primary crystal domain is the exclusive source of the measured magnetostriction signal. While we can not rule out the presence of secondary twinned crystals in our sample, these defects could not possibly cause such large differences as $\approx 20$ \% in the peak heights of our main Landau spectrum, if they are otherwise undetectable.

To summarize, our magnetostriction data finds no measureable difference between the valleys regarding their total occupation, but a sizeable difference in the DOS at the Fermi level. At first sight, this appears paradoxical. But in each Landau tube, states at the Fermi level only constitute a small fraction of all carriers. Apparently the experimentally observed valley imbalance only affects these carriers.

Having ruled out the evident possibilities of strain and secondary twinned crystals as causes of the observed symmetry breaking, interactions among electrons could be considered responsible for the valley-dependent density of states. A suppressed DOS at the Fermi level is frequently observed in correlated systems. One well-known example is the Coulomb pseudogap in disordered semiconductors~\cite{Efros75}, where long-range interactions between electrons lead to a soft depletion in the density of localized states at the Fermi level. For such a Coulomb-gap to become valley-selective, the isotropic Coulomb interaction would have to be combined with a source of anisotropy, like an anisotropic distribution of defects in the bismuth crystal, offering electrons an unequal number of localized states along the symmetry axes. While no real crystal has an exactly equal defect density along all crystalline axes, we can not quantify the exact difference for our sample.

Another well-documented source of anisotropy is the effective mass itself, which for electrons in bismuth differs by a factor of 200 along the binary and bisectrix axis, respectively~\cite{Dresselhaus71}. Due to this extreme mass anisotropy, Abanin {\it et al.}~\cite{Abanin10} already suspected that valley nematicity - a theoretical scenario proposed for multiple-valley quantum Hall systems with significant mass anisotropy such as AlAs - might be relevant for bismuth, too. Valley nematicity occurs when electrons reduce the rotational symmetry of the crystal by unequally occupying equivalent valleys. According to Abanin \emph{et al.}~\cite{Abanin10}, if the mass anisotropy in a two-dimensional system is large enough, the exchange energy will favor such an inequality. Zhu {\it et al.}~\cite{Zhu11a} suggested this scenario as a possible explanation for the symmetry breaking they observed in the magnetoconductivity, which was met by the objection that the theory developed for quantum Hall systems refers explicitly to a two-dimensional system without the kinetic energy. How can it be applied to bulk bismuth, in which electrons keep a finite energy even in the presence of Landau quantization? Our results provide a possible answer to this theoretical objection as well: the imbalance between the two valleys emerges only when a LL is evacuated, which is when the kinetic energy of the electrons along the $B$ direction vanishes (center of the parabolic dispersion sketched in Fig.~\ref{Fig1}b). In the absence of kinetic energy, theoretical considerations on the energetics of valley nematicity in a two-dimensional system~\cite{Abanin10} become relevant to bulk bismuth as well.

Our context of investigation is different from both disordered semiconductors and multiple-valley quantum Hall systems, but our experimental observation points to a modification of the fine structure of the DOS at the Fermi level, presumably by interaction, which is not identical between the three valleys. This frames a specific challenge to theory. The link between electron correlation and valley imbalance in the specific case of $Z_{3}$ symmetry in a three-dimensional context begs a thorough theoretical investigation.

\section*{Acknowledgement}
We would like to thank C. Geibel, P. Gegenwart, J. Alicea, L. Balents, M. Doerr, B. Fauqu\'e, and Y. Fuseya for valuable discussions, as well as T. L\"{u}hmann and N. Oeschler for technical and organizational support. R. K. is supported by the DFG Research Unit 960 (Quantum Phase Transitions). KB is supported by ANR through the QUANTHERM project and acknowleges the hospitality of the Aspen Center for Physics.\\\\
Correspondence to:\\\\
Robert K\"uchler$^{1,\dagger}$ e-mail: kuechler@cpfs.mpg.de\\
Lucia Steinke$^{1,\dagger}$ e-mail: lucia@drs-steinke.de\\

\end{document}